\documentclass[12pt]{article}
\usepackage{amsmath,amssymb}
\newcommand{\Ft}{F_{T}}
\newcommand{\Zparity}{\ensuremath{\mathbb{Z}_2}}
\newcommand{\BLambda}{\boldsymbol{\lambda}}

\begin{document}
\topmargin -1.0cm
\oddsidemargin -0.8cm
\evensidemargin -0.8cm
\pagestyle{empty}
\begin{flushright}
UAB-FT-518\\
June 2001
\end{flushright}
\vspace*{5mm}

\begin{center}

{\Large\bf 
Supersymmetric theories with compact extra dimensions}\\ 
\vspace*{5mm}
{\Large\bf 
in $N=1$   superfields}\\
\vspace{2.0cm}

{\large Daniel Mart{\'\i} and  Alex Pomarol}\\

\vspace{.6cm}
{\it IFAE, Universitat Aut{\`o}noma de Barcelona, 
E-08193 Barcelona, Spain}
\vspace{.4cm}
\end{center}

\vspace{1cm}
\begin{abstract}
We present an $N=1$ superfield formulation 
of   supersymmetric gauge theories 
with a compact  extra dimension. 
The formulation 
incorporates
the radion superfield and allows to
write
supersymmetric theories
on warped gravitational backgrounds.
We apply it to study the breaking of supersymmetry
by the $F$-term of the radion, and show that,
for  flat extra dimensions,
this leads to 
the same mass spectrum as 
in  Scherk-Schwarz  models of supersymmetry breaking.
We also consider scenarios where supersymmetry
is broken on a boundary of a warped extra dimension
and compare them with anomaly mediated models.
\end{abstract}

\vfill
\begin{flushleft}
\end{flushleft}
\eject
\pagestyle{empty}
\setcounter{page}{1}
\setcounter{footnote}{0}
\pagestyle{plain}


\section{Introduction}

Supersymmetry and extra dimensions are well-motivated extensions
of the Standard Model.
They could play a role in
 the hierarchy problem, or be crucial ingredients 
in a quantum description of gravity, as is
the case in  string theory. 
Also extra dimensions and supersymmetry could be connected to the origin
of the  electroweak symmetry breaking \cite{ewsb}.

To study  supersymmetric theories it is very useful
to have a  superfield description
where supersymmetry invariance is manifest,
and nonrenormalization theorems are easily derived \cite{wb}.
The $N=1$ superfield formalism has been extensively analyzed in 
four dimensions.
In  higher dimensions, however,
supersymmetry is usually presented in component fields,
since a superfield description is usually not known.
A first attempt to write  higher-dimensional
supersymmetric theories in superfields
was presented in Ref.~\cite{mss} for  theories  in 10 dimensions,
and  
has  been  recently extended to other dimensions in Ref.~\cite{agw}.
The formulation of Refs.~\cite{mss,agw}   is based  on writing
higher-dimensional supersymmetric theories using  $N=1$ 
four-dimensional superfields.
These are the ordinary superfields defined in a  4D superspace. 
Higher-dimensional supersymmetric theories contain the 4D supersymmetry
and therefore it is always possible to write them 
using $N=1$ superfields.
With this formulation, only the $N=1$ supersymmetry is manifest.
In spite 
of this limitation, we think that the formulation is
already very useful for supersymmetric theories with extra dimensions.
The effective action is  simpler to write than  in component fields,
and the bulk-boundary couplings are easily obtained.
Nonrenormalization theorems can also be derived.

Here we will use the $N=1$ superfield formulation to
write the action of a 5D supersymmetric theory with 
a compact extra dimension.
The important new ingredient  of our formulation is that 
it incorporates the radion superfield $T$.
This will allow us   to write the supersymmetric action for fields
living in either  a flat or  a warped extra dimension.
In particular, we will consider a gauge theory with 
 5D vector multiplets and charged hypermultiplets in an
extra dimension 
(1) flat and  compactified 
in a circle $S^1$ or orbifold $S^1/\Zparity$, 
(2) warped as in the Randall-Sundrum (RS) scenario \cite{rs}.

We will apply this superfield formulation to
study supersymmetry  breaking induced by the  $F$-term of the 
radion superfield. 
This occurs when a constant superpotential
is present in the bulk of the extra dimension.
For a flat extra dimension
we will show that 
this is equivalent to break  supersymmetry by boundary conditions
(Scherk-Schwarz (SS) mechanism \cite{ss}).
In particular, we will derive the  mass spectrum of the models of 
Refs.~\cite{pq,bhn}.
Our formulation therefore will provide a superfield
description of 
the SS mechanism and will show that 
the SS  breaking of supersymmetry is spontaneous.

For a warped extra dimension, we will consider the 
breaking of supersymmetry induced by a  boundary superpotential,
and derive the mass spectrum in the gauge sector.
We will show the similarities of this  breaking
with
 models of anomaly mediated supersymmetry breaking (AMSB) 
\cite{amsb}.

A final comment is in order.
In our formulation we will be only considering part 
of the 5D supergravity multiplet, 
the radion superfield $T$. 
Therefore our approach does not incorporate the full
5D gravitational sector.
The action derived below
must be considered as that of 
supersymmetric theories on  nontrivial gravitational backgrounds.
A complete formulation with the full 5D supergravity multiplet is
a subject of future research.

\section{5D superfield action with  a flat and compact extra dimension}

Let us  consider a 5D theory  in ${\cal M}^4\times S^1$.
The  metric is given by
\begin{equation}
    ds^2 = \eta_{\mu\nu} dx^\mu dx^\nu + R^2 dy^2 \, ,
\label{flat}
\end{equation}
where $R$ is the radion of the extra dimension labeled by $y$, which
ranges from 0 to $2\pi$.  We want to derive the action of 
superfields living on the 5D gravitational background of
Eq.~(\ref{flat}).
For this purpose, we need to promote $R$ to a superfield.  
This
corresponds to a 4D chiral superfield $T$ that, together with $R$, it
is known to contain the fifth-component of the graviphoton $B_5$, the
fifth-component of the right-handed gravitino $\Psi_R^5$
and a complex auxiliary field $F_T$, and we will write it as
\begin{equation}
 T=  R + i B_{5} + \theta \Psi^5_R + \theta^{2} \Ft\, .
\end{equation}

\subsection{Vector supermultiplet}
\label{sec:gauge}

The off-shell 5D $N=1$ vector supermultiplet 
consists of  a 5D vector $A_M$, two
Weyl gauginos $\lambda_{1,2}$, a real scalar $\Sigma$,
and a real and complex  auxiliary field $D$ and  $F_\chi$ respectively.
Under the
$N=1$ supersymmetry, they form a vector supermultiplet $V$ and a chiral
supermultiplet ${ \chi}$
\footnote{We will follow the notation of  Ref.~\cite{wb}.}:
\begin{align}
    V & = -\theta \sigma^{\mu} \bar{\theta} A_{\mu} -
    i\bar{\theta}^2\mspace{-2mu}\theta\lambda_{1} +
    i\theta^2\mspace{-2mu}\bar{\theta}\bar{\lambda}_{1} +
    \frac{1}{2} \bar{\theta}^2\mspace{-2mu}\theta^2 D\, ,\nonumber \\
    \chi &= \frac{1}{\sqrt{2}} \left(\Sigma + i A_{5}\right) +
    \sqrt{2} \theta \lambda_{2} + \theta^{2} F_\chi\, ,
\end{align}
where $V$ is given in the Wess-Zumino gauge.
Let us first consider an Abelian theory.  
Under a gauge transformation, the superfields transform as
\begin{align}
V&\rightarrow V+\Lambda+\Lambda^{\dag}\, ,\nonumber\\
\chi&\rightarrow \chi+\sqrt{2}\partial_5\Lambda\, ,
\end{align}
where $\Lambda$ is an arbitrary chiral field. 
The gauge invariant action is
given by
\begin{equation}
    S_5 = \int d^{5}\! x\, 
\left[ \frac{1}{4g^2_5} \int d^2\mspace{-2mu}\theta\, T
        W^{\alpha}W_{\alpha} + \text{h.c.} + \frac{2}{g^2_5} \int
         d^4\mspace{-2mu}\theta\: \frac{1}{(T +
          T^{\dag})} \left( \partial_5 V - \frac{1}{\sqrt{2}} (\chi +
            \chi^{\dag})\right)^2 \right]\, .
\label{GaugeAb}
\end{equation}
It is easy to check that this superfield action leads to the right
action in component fields.  We must perform the integrals over $\theta$
taking $\langle T\rangle=R$, eliminate the auxiliary fields using their
equation of motion
\begin{equation}
F_\chi=0\, , \qquad\qquad D= - \frac{\partial_{5}\Sigma}{R^2}\, ,
\end{equation}
and rescale the fields according to $\Sigma
\rightarrow R \Sigma$, $ \lambda_2
\rightarrow - i R \lambda_2$.  We finally obtain
\begin{equation}
S_5 = \frac{1}{g^2_5}\int d^{5}\! x\, \sqrt{-g} \biggl[ 
-\frac{1}{2}\partial_{M}\Sigma
    \partial^{M} \Sigma - \frac{1}{4} F_{MN} F^{MN} 
+ \frac{i}{2} \bar{\BLambda}_{i}
    \gamma^{M} \partial_{M} {\BLambda}_{i}
    \biggr]\, ,
    \label{S_abelian}
\end{equation}
where we have defined the symplectic-Majorana spinors 
$\left[{\BLambda}_{i}\right]^{T} \equiv (\lambda_{i},
\epsilon_{ij}\bar{\lambda}_{j})$ in order to
make Eq.~(\ref{S_abelian}) manifestly invariant under the $SU(2)$
automorphism group \cite{pm}.

For the non-Abelian case, the second term of Eq.~(\ref{GaugeAb}) must be
replaced by
\begin{equation}
   \frac{2}{g^2_5} \int d^4\mspace{-2mu}\theta\: 
\frac{1}{(T+ T^{\dag})} {\rm Tr} \Bigl[
        \{e^{V/2}, \partial_5 e^{-V/2}\} + \frac{1}{\sqrt{2}} (e^{V/2}
        \chi^{\dag} e^{-V/2} + e^{-V/2} \chi e^{V/2}) \Bigr]^2\, ,
\label{nonabelian}
\end{equation}
that is gauge invariant under the gauge transformation
\begin{align}
    \chi & \rightarrow U^{-1} (\chi - \sqrt{2} \partial_5) U\, , & e^{V} &
    \rightarrow U^{-1}e^{V} U^{-1\, \dagger}\, , \\
    \chi^{\dag} & \rightarrow U^\dagger (\chi^{\dag} + \sqrt{2} \partial_5)
    U^{-1\, \dagger}\, ,& e^{-V} & \rightarrow U^\dagger e^{-V} U\, ,
\label{nona}
\end{align}
where $U = e^{-\Lambda}$, $U^\dagger = e^{-{\Lambda}^{\dag}}$,
$\Lambda=\Lambda^{a}T^{a}$, $\chi \equiv \chi^{a}T^{a}$ and $V\equiv V^{a}
T^{a}$.  For $T=\text{constant}$, Eq.~(\ref{nonabelian})   differs from 
Ref.~\cite{agw} only
in chiral or antichiral terms which vanish under the integration over
the whole superspace $\int\! d^4\mspace{-2mu}\theta$.  Nevertheless, these
terms are nonzero for the case in which $T$ is a superfield with a
nontrivial $\theta$ dependence, and they must be taken into account.

\subsection{Hypermultiplet}

The off-shell 5D hypermultiplet consists in two complex scalars, $\phi$ and
$\phi^c$,  a Dirac fermion $\Psi$ and two complex
auxiliary fields $F_\Phi$ and 
$F_{\Phi^c}$. It can be arranged in two $N=1$
chiral superfields, $\Phi$ and $\Phi^c$. Assuming that they are charged
under some gauge group and transform as $\Phi\rightarrow U^{-1}\Phi$ and
$\Phi^c\rightarrow \Phi^c U$, we have that the 5D action for the
hypermultiplet is given by
\begin{equation}
    S_5 = \int d^{5}\! x\, \left\{ \int d^4\mspace{-2mu}\theta\: \frac{1}{2}
    (T + T^{\dag} ) \,\left(\Phi^{\dag} e^{-V} \Phi + 
\Phi^{c} e^{V} \Phi^{c\, \dag}\right) + \int
    d^2\mspace{-2mu} \theta\, \Phi^{c} \Bigl[
    \partial_{5}-\frac{1}{\sqrt{2}}\chi\Bigr] \Phi +  
\text{h.c.} \right\}\, .
\label{eq:hyperS}
\end{equation}
For $\langle T\rangle=R$, the auxiliary fields 
are given by
\begin{align}
    F_\Phi & =\frac{1}{R}\left(\partial_{5}+ \frac{1}{2}(\Sigma -
        i A_5) \right)\phi^{c\,\dag}\, ,
      &   F_{\chi}^{a} & = \frac{g^2_5R}{\sqrt{2}} \,\phi^{\dag} T^{a}
        \phi^{c\,\dag}\, ,\nonumber\\
      F_{\Phi^{c}}^{\dag} & = - \frac{1}{R}\left( \partial_{5}-
          \frac{1}{2}(\Sigma + i A_5) \right)\phi\, ,
      \label{eq:relF} &  D^{a} & = -\frac{\partial_{5}\Sigma^a}{R^2} +
      \frac{g_{5}^2}{2}(\phi^{\dag}T^{a}\phi - \phi^{c} T^{a}
      \phi^{c\,\dag})\, .
\end{align}
By rescaling
 $\Sigma$ and $\lambda_2$ as in Section \ref{sec:gauge},
Eq.~(\ref{eq:hyperS}) gives
\begin{eqnarray}
        S_5 &=& \int d^{5}\! x \sqrt{-g} \, \Bigl\{  - | D_{M} \phi
        |^2 - |D_{M} \phi^{c}|^2
        + i \bar{\Psi} \gamma^{M} D_{M} \Psi
               -\frac{1}{\sqrt{2}} 
(\phi^{c} \bar{\BLambda}_{1} \Psi - \phi^{\dag} \bar{\BLambda}_{2} \Psi)
        + \text {h.c.} \nonumber\\
        & -&\frac{i}{2}  \bar{\Psi} \Sigma \Psi - 
\frac{1}{4}(\phi^{\dag}_i \Sigma^2 \phi_i)
        - \frac{g^2_5}{8}
        \sum_{m,a}(\phi_{i}^{\dag} (\sigma^{m})_{ij} T^{a} \phi_{j} )^2
        \Bigr\}\, ,
\label{hyperfinal}
\end{eqnarray}
where in the last two terms we have defined
$\{\phi_1,\phi_2\} \equiv \{\phi,\phi^{c\,\dag}\}$ and the
gauge covariant derivative as
$D_M = \partial_M - \frac{i}{2}  A_M^{a} T^{a}$.
A supersymmetric mass for the hypermultiplet, 
\begin{equation}
\int d^2\mspace{-2mu} \theta\, \Phi^c m \Phi+\text{h.c.}\, ,
\label{susymass}
\end{equation}
can be easily included 
by performing in Eq.~(\ref{eq:hyperS})
the shift 
$\chi\rightarrow \chi-\sqrt{2}m$
that in Eq.~(\ref{hyperfinal}) corresponds to 
$\Sigma\rightarrow \Sigma-2m$.

\subsection{The orbifold $S^1/\Zparity$ and bulk-boundary couplings}

The above bulk action is not modified if the extra dimension is
compactified in the orbifold $S^1/\Zparity$, which corresponds to the
circle $S^1$ with the identification $y\leftrightarrow -y$.  This
identification leads to a manifold with two boundaries at $y=0$ and at
$y=\pi$.

There can be fields living on these 4D boundaries. They respect an $N=1$
supersymmetry and therefore their superfield action is the ordinary one.
The couplings of the boundary fields to the bulk fields
are easily obtained using superfields. Assuming
that $V$ and $\chi$ are respectively even and odd under the $\Zparity$ parity,
 we have that $\chi$ vanishes 
on the boundaries and therefore  only $V$ 
couples to fields on the boundaries.  For the
hypermultiplet, if we assume that $\Phi$ and $\Phi^c$ are respectively
even and odd under the $\Zparity$, we have that only $\Phi$ can couple
to the boundary fields.  For a chiral superfield $Q$ living on the $y=0$
boundary
these couplings are simply given by
\begin{equation}
S_5=\int d^{5}\! x\, 
\Bigg[\int d^4\mspace{-2mu}\theta\,\Big(Q^{\dag} e^{-V} Q+e^{-V}\xi\Big) 
+\int d^2\mspace{-2mu}\theta\,
 W(\Phi,Q)+ \text{h.c.}\Bigg]\delta(y) \, ,
\label{bulkbound}
\end{equation}
where $W$ is a superpotential that can depend on 
$\Phi$ and $Q$, and $\xi$
is a Fayet-Iliopoulos term that can be present for an Abelian 
vector supermultiplet.  
The boundary couplings Eq.~(\ref{bulkbound})
change the auxiliary field equation of
motion by $\delta$-function terms:
\begin{align}
D &=  -\frac{\partial_{5} \Sigma}{R^2} + \frac{g^2_5}{2} \delta(y)
\bigl(Q^{\dag} Q + \xi \bigr)\, , \nonumber\\
F_{\Phi} & =\frac{1}{R}\left(\partial_{5}+
            \frac{1}{2}(\Sigma - i A_5) \right)\phi^{c\, \dag}
        - \delta(y) \left.
\frac{\partial W}{\partial \Phi}\right\vert_{\Phi=\phi}\, .
\end{align}
Similarly we can obtain the couplings on the boundary at $y=\pi$. 

\section{Superfield action in a warped 5D space}

The above action can be generalized to the case where the extra
dimension is warped. As an example, we will consider the 
RS scenario \cite{rs} where the extra dimension 
$y$ is compactified on an orbifold $S^1/\Zparity$ of radius $R$, with 
$-\pi \leq y \leq \pi $.
The 5D space is defined by the metric
\begin{equation}
ds^2=e^{-2R\sigma}\eta_{\mu\nu}dx^\mu dx^\nu+R^2dy^2\, ,
\label{rsmetric}
\end{equation}
where
\begin{equation}
\sigma=k|y|\, ,
\end{equation}
and $1/k$ is the curvature radius.  This space corresponds to a slice of
AdS$_5$.  The supersymmetric version of the RS model
has been recently studied in Refs.~\cite{gp,rssusy}.
The supersymmetric conditions for vector and hypermultiplets 
on the background of Eq.~(\ref{rsmetric})
were derived
in Ref.~\cite{gp}.  
Here we will present the action written in
superfields.

\subsection{Vector supermultiplet}

The action for the Abelian vector superfield is given by
\begin{equation}
   S_5 = \int d^{5}\! x\, \left[ \frac{1}{4g^2_5} \int 
d^2\mspace{-2mu}\theta\, T
        W^{\alpha}W_{\alpha} + \text{h.c.} + \frac{2}{g^2_5} \int
        d^4\mspace{-2mu}\theta\,
 \frac{e^{-(T + T^{\dag})\sigma}}{(T +
          T^{\dag})} \left( \partial_5 V - \frac{1}{\sqrt{2}} (\chi +
            \chi^{\dag})\right)^2 \right]\, .
\label{gaugew}
\end{equation}
The  auxiliary fields are  given by
\begin{equation} 
F_\chi=0\, , \qquad\qquad D= - \frac{e^{-2R\sigma}}{R^2} \left(\partial_{5} - 
     2R\sigma' \right) \Sigma\, ,
\end{equation}
where $\sigma'=\partial_5\sigma$.
After the rescaling $\Sigma \rightarrow R \Sigma$, $\lambda_1
 \rightarrow e^{-3R\sigma/2} \lambda_1$, $\lambda_2 \rightarrow - i R
 e^{-R\sigma/2} \lambda_2$, we obtain the action
\begin{equation}
    S_5 = - \frac{1}{g^2_5} \int d^{5}\! x\, \sqrt{-g} \Biggl[ \frac{1}{2}
    \bigl(\partial_{M}\Sigma \bigr)^2 + \frac{1}{2} m^{2}_{\Sigma} \Sigma^{2} +
    \frac{1}{4} F_{MN}^2 - \frac{i}{2} \bar{\BLambda}_{i}
    \gamma^{M} D_{M} {\BLambda}_{i} -
m_{\lambda}\frac{i}{2}\bar{\BLambda}_{i} [\sigma_{3}]_{ij}
    {\BLambda}_{j} \Biggr]\, ,
\end{equation}
where $\sigma_3=\text{diag}(1,-1)$, $D_M {\BLambda}_{i} = \partial_M
{\BLambda}_{i} + \Gamma_M [\sigma_{3}]_{ij} {\BLambda}_{j}$, being $\Gamma_M$
the spin connection, $\Gamma_M =( \sigma' \gamma_{5}\gamma_{\mu} /2,0)$,
and where 
\begin{equation}
    m^2_{\Sigma} = -4 k^2 + 2\frac{\sigma''}{R}\, ,\qquad
    m_{\lambda} = \frac{1}{2}\sigma'\, ,
\end{equation}
in agreement with Ref.~\cite{gp}.

\subsection{Hypermultiplet}

The action is given by
\begin{align}
    S_5 =& \int d^{5}\! x\, \left\{ \int d^4\mspace{-2mu}\theta \:
        \frac{1}{2} (T + T^{\dag} ) e^{-(T+ T^{\dag})\sigma}
        \,\bigl(\Phi^{\dag} e^{-V}\Phi + \Phi^{c} e^{V}\Phi^{c\,\dag}\bigr) \right.\nonumber\\
    & + \left.\int d^2\mspace{-2mu} \theta\, e^{-3 T \sigma}\, \Phi^c
        \Bigl[ \partial_{5} -\frac{1}{\sqrt{2}}\chi - \Bigl( \frac{3}{2}
        - c \Bigr) T\sigma^\prime \Bigr] \Phi + \text{h.c.}
    \right\}\, ,
\label{hyperw}
\end{align}
where following Ref.~\cite{gp} we have parametrized the hypermultiplet
mass as $c\sigma^\prime$.  
Neglecting for simplicity the gauge sector, 
we have that 
the  auxiliary fields are given by
\begin{align}
    F_{\Phi} = & \frac{e^{-R\sigma}}{R} \Bigl[\partial_{5} +
    \Bigl(\frac{3}{2} - c \Bigr)R \sigma' \Bigr] \phi^{c\, \dagger}
\, ,\nonumber\\
    F_{\Phi^c}^\dagger = & - \frac{e^{-R\sigma}}{R} \Bigl[\partial_{5} -
    \Bigl(\frac{3}{2} - c \Bigr)R \sigma' \Bigr] \phi\, ,
\end{align}
that after the rescaling $\Psi \rightarrow e^{-R\sigma/2} \Psi$ yields,
\begin{equation}
    S_5 = \int d^{5}\! x \sqrt{-g} \, \Bigl\{ \bigl( - |\partial_{M}
    \phi|^2 - |\partial_{M}  \phi^{c}|^2 - m_{\phi}^2 |\phi|^2 -
m_{\phi^c}^2     |\phi^{c}|^2  + i \bar{\Psi} \gamma^{M}(\partial_{M} +
    \Gamma_{M}) \Psi - i m_{\Psi} \bar{\Psi}\Psi \bigr)\Bigr\}\, ,
\end{equation}
where
\begin{equation}
    m^2_{\phi,\phi^c}  = \Bigl( c^2 \pm c - \frac{15}{4} \Bigr)k^2 +
    \Bigl(\frac{3}{2} \mp
    c\Bigr) \frac{\sigma''}{R}\, ,\qquad    m_{\Psi}  = c\sigma'\, .
\end{equation}

\subsection{Low-energy 4D effective theory}

At energies below the KK masses (that in the warped case are of order 
$ke^{-Rk\pi}$), the KK can be integrated out and 
the effective theory can be written with 
only the massless sector.
For  the 5D vector superfield this corresponds to the 
zero mode of the    superfield $V$.
Its wave-function $f_0(y)$ is determined by $\partial_5 f_0=0$, so
it is  $y$-independent \cite{gaugewarped}.  
The 4D effective lagrangian for 
the massless mode of $V$
is therefore the same as the one
with a flat extra dimension:
\begin{equation}
{\cal L}_{4D}=\frac{\pi}{2g^2_5} \int d^2\mspace{-2mu}\theta\, T
        W^{\alpha}W_{\alpha} + \text{h.c.}\, . 
\label{vector0}
\end{equation}
For the hypermultiplet
only the chiral superfield $\Phi$ (even under $\Zparity$)
has a massless  mode.
Its wave-function satisfies
$ [\partial_5-(\frac{3}{2} - c)T \sigma' ] f_0=0$
that yields
$
f_0=e^{(\frac{3}{2} - c ) T \sigma}
$.
After integrating over $y$, we get 
the effective lagrangian 
\begin{equation}
{\cal L}_{4D}= \int d^4\mspace{-2mu}\theta
    \: \frac{1}{\left(\frac{1}{2}-c\right)k}\left(
e^{(\frac{1}{2}-c)(T+T^\dagger)k\pi}-1\right) 
\Phi^\dagger e^{-V}\Phi\, .
\label{hyper0}
\end{equation}

\subsection{Bulk-boundary couplings}

For a chiral superfield $Q$ living on the $y=0$ boundary the
bulk-boundary couplings are the same as those in Eq.~(\ref{bulkbound}).
On the other boundary, at $y=\pi$, we have
\begin{equation}
S_5=\int d^{5}\! x\, 
\Bigg[\int d^4\mspace{-2mu}\theta\, e^{-(T+ T^\dagger)k\pi} 
\Big( Q^\dagger e^{-V} Q+e^{-V}\xi\Big) 
+\int d^2\mspace{-2mu}\theta\, e^{-3Tk\pi} 
W(\Phi,Q)+ \text{h.c.}
\Bigg]\delta(y-\pi)\, .
\label{boundtev}
\end{equation}

\section{Supersymmetry breaking by the radion $F$-term}

As an application of the action derived above, we will calculate the 
spectrum of soft masses that is obtained when supersymmetry is broken
by a nonzero  $F$-term  of the radion superfield.
We will first consider the case of a flat extra dimension
with a constant ($y$-independent)
  $F_T$.
We will show that the resulting mass 
spectrum is the same as that in models with  SS supersymmetry breaking.
For a 
warped extra dimension we will consider the case when
 $F_T$ is induced on the boundary. 
We will see that 
$F_T$ is
 exponentially suppressed,
generating  small soft masses.

\subsection{A flat extra dimension}

The simplest model to break supersymmetry 
by the radion $F$-term is the no-scale model \cite{noscale}.
In 
superfields this is given by
\begin{equation}
S_5=\int d^{5}\! x\, \left[-3
M^3_5\int d^4\mspace{-2mu}\theta\,
 \varphi^\dagger\varphi\frac{(T+T^\dagger)}{2}+
\int d^2\mspace{-2mu}\theta\, \varphi^3 W+\text{h.c.}\right]\, ,
\label{radiusa}
\end{equation}
where $W$ is a  ``superpotential'' 
that can arise in gauged supergravity
theories \cite{W}. 
We are considering that  $W$ is constant ($y$-independent) and real.
Eq.~(\ref{radiusa}) corresponds to the effective action of the radion
of  a flat extra dimension.
We
have introduced  
the conformal compensator $\varphi$ \cite{compensator}. 
This is a non-propagating superfield
useful to incorporate the supergravity effects to the effective 
radion potential. 
The superfield  $\varphi$ makes the 4D supergravity action manifestly 
invariant under a conformal transformation.
The breaking of  the superconformal group down to the 
super-Poincare group is parametrized by the scalar component of $\varphi$
and the  $F$-term of $\varphi$ corresponds to
the auxiliary component of the supergravity multiplet:
\begin{equation}
\varphi=1+\theta^2 F_\varphi\, .
\end{equation}
{}From Eq.~(\ref{radiusa}) we obtain
\begin{equation}
F_T=\frac{2W}{M^3_5},\qquad\qquad  F_\varphi=0\, ,
\label{fterms}
\end{equation}
that leads to a vanishing potential for the radion (at tree level).
Therefore if the vacuum expectation value 
of $W$ is nonzero, supersymmetry is broken 
by a nonzero $F_T$ with zero cosmological constant.
This is  a very simple example of breaking supersymmetry
by $F_T$. 
This model, however, does not stabilize  the radius $R$. 
For this purpose,
one could need some extra massive matter fields, 
as for example in Ref.~\cite{pop}.
Other scenarios have been proposed in 
Ref.~\cite{ls}.
In what follows, we will assume 
that the radius is stabilized without affecting Eq.~(\ref{fterms}).

Let us derive the soft masses in the gauge sector
for this scenario.
Taking $T=R+\theta^2 F_T$
in Eq.~(\ref{GaugeAb}),  
we obtain an extra mass term for the 5D gauginos:
\begin{equation}
{\cal L}_{soft}=\frac{1}{2}\frac{\Ft}{2R}\:
    {\BLambda}_{i}^{\mspace{2mu}T}\! C {\BLambda}_{i}\, ,
\label{majorana}
\end{equation}
where $C$ is the charge-conjugation matrix in 5D.
Eq.~(\ref{majorana}) is a Majorana mass term.
In the case of the orbifold $S^1/\Zparity$
where 
we  decompose the 5D fields in KK states, 
$\lambda_1=\sum\cos(ny) \lambda^{(n)}_1(x)$ and
$\lambda_2=\sum\sin(ny) \lambda^{(n)}_2(x)$, 
the gaugino masses are given by
\begin{equation}
{\cal L}_{mass}=\frac{F_T}{2R}
\lambda_1^{(0)}\lambda_1^{(0)}+
\frac{1}{R}\sum_{n=1}^\infty
\begin{pmatrix}
    \lambda_{1}^{(n)} & \lambda_2^{(n)}
\end{pmatrix}
\begin{pmatrix}
    \Ft/2 & n \\
    n & \Ft/2
\end{pmatrix}
\begin{pmatrix}
    \lambda_1^{(n)}\\
    \lambda_2^{(n)}
\end{pmatrix}+\text{h.c.}\, ,
\end{equation}
which give rise to Majorana gaugino  masses $|\Ft/2\pm n|/R$.
This is the same spectrum as the one obtained by SS supersymmetry breaking
with an  $R$-symmetry \cite{pq}.
The correspondence is $F_T/2=q_R$ that implies $W/M^3_5=q_R$,  
where  $q_R$ is the $R$-charge defined in Ref.~\cite{pq}. 
For $F_T=1$
it also corresponds to the gaugino mass spectrum of Ref.~\cite{bhn}.
Notice that the scalar $\Sigma$ does not get a soft mass
as it is also predicted by SS supersymmetry breaking since the $R$-charge
of $\Sigma$ is zero.

For the hypermultiplet 
a nonzero $F_T$ gives in Eq.~(\ref{eq:hyperS}) 
an extra
contribution to the $F$-terms of the scalars.
These are now given  by 
\begin{eqnarray}
       && F_{\Phi}= \frac{1}{R}\left([\partial_{5}+i\alpha]\phi^{c\,\dagger} -
            \frac{1}{2}\Ft \phi\right)\, ,\nonumber\\
 && F_{\Phi^{c}}^\dagger =-\frac{1}{R}\left([\partial_{5}+i\alpha]{\phi}+
    \frac{1}{2}\Ft \phi^{c\, \dagger}\right)\, ,
\label{eq:relFb}
\end{eqnarray}
where, for later use,  we have also included a supersymmetric mass,
Eq.~(\ref{susymass}), with $m=i\alpha$.
After KK reduction in a circle $S^1$ (imposing periodic boundary 
conditions), 
the mass term  for 
the scalars is given by 
\begin{equation}   
{\cal L}_{mass}=-\frac{1}{R^2}\sum^{\infty}_{n=-\infty}
\begin{pmatrix}
       \phi^{(n)\, \dagger}&\phi^{c\, (n)\, \dagger}
    \end{pmatrix}
\begin{pmatrix}
       (n+ \alpha)^2+ F^2_T/4
 &        i(n+\alpha)F_T  \\
        -i(n+\alpha)F_T 
 &       (n+ \alpha)^2+ F^2_T/4
    \end{pmatrix}
\begin{pmatrix}
       \phi^{(n)} \\
         \phi^{c\, (n)}
    \end{pmatrix}\, ,
\label{massonehyper}
\end{equation}
that  corresponds to scalars with  masses $|n+\alpha\pm F_T/2|/R$.
Fermions do not get soft masses 
and the  mass spectrum  is given by $|n+\alpha|/R$.
Let us compare the mass spectrum of Eq.~(\ref{massonehyper})
with that
in the models of Refs.~\cite{pq} and \cite{bhn}.

In Ref.~\cite{bhn} 
quarks and leptons were associated to  hypermultiplets
that would correspond to the above hypermultiplet
with $\alpha=0$, $F_T=1$ and, after orbifolding,
$n=0,1,2,...$.
To derive the mass spectrum of the Higgs of Ref.~\cite{bhn},
 we must proceed 
as follows.
Let us 
absorb the supersymmetric mass $m=i\alpha$  of the hypermultiplet
by the redefinition
\footnote{This $y$-dependence of the hypermultiplet 
can be the result of imposing the 
(supersymmetric) boundary condition 
$\Phi^{'}(x,y+2\pi) = e^{i 2\pi\alpha }\Phi^{'}(x,y)$
and similarly for
$\Phi^{c\, '}$.} 
\begin{equation}
\Phi^{'} = e^{i \alpha y}\Phi
\, ,\qquad 
\Phi^{c\, '} 
= e^{-i \alpha y}\Phi^c\, ,
\end{equation}
and assign  the following $\Zparity$
parities:
$\Phi^{'}(y)\rightarrow\Phi^{'}(-y)$
and $\Phi^{c\, '}(y)\rightarrow-\Phi^{c\, '}(-y)$.
For $\alpha=1/2$, we have
 that the KK decomposition 
in  the $S^1/\Zparity$ orbifold 
is given by
$\Phi^{'} = \sum_n \cos[(n+1/2) y]\Phi^{(n)}(x)$
and $\Phi^{c\, '}= \sum_n \sin[(n+1/2) y]\Phi^{c\, (n)}(x)$ 
where $n=0,1,2,...$.
The mass spectrum  is supersymmetric and is given by $|n+1/2|/R$. 
Let us now turn on the $F_T$ and break supersymmetry. 
The scalar mass matrix will be given by 
Eq.~(\ref{massonehyper}) with $\alpha=1/2$ and $n=0,1,2,...$,
and therefore the scalar
masses will be $|n+1/2\pm F_T/2|/R$. 
For $F_T=1$ we obtain the same mass spectrum as that of the
Higgs in Ref.~\cite{bhn}.
There is a single massless scalar that is associated 
with the SM Higgs.

In Ref.~\cite{pq} 
quarks and leptons were localized on the boundaries
of the orbifold
while the Higgs sector was living in the bulk.
To give  masses to the fermions, 
the Higgs sector had to consist in two
hypermultiplets
$(\Phi_i, \Phi^{c}_i)$ $i=1,2$.
Assigning the $\Zparity$ parity as
$\Phi(y) \rightarrow \eta \Phi(-y)$, with
$\eta=+1 (-1)$ for $\Phi_{1}$ and $\Phi_{2}^{c}$ (
$\Phi^{c}_{1}$ and $\Phi_{2}$), 
the Higgs sector
 can be written as
\begin{equation}
        S_5  = \int d^{5}\! x\,
         \biggl\{ \int
                d^4\mspace{-2mu}\theta\: \frac{1}{2} (T + T^{\dag} ) \,
                \left( \Phi^{\dag}_i \Phi_i + \Phi^{c}_i \Phi^{c\, \dag}_i
                \right) + \int d^2\mspace{-2mu} \theta\, \Phi^{c}_i
                (\partial_5\delta_{ij}+m\epsilon_{ij}) \Phi_j +\text{h.c.}
            \biggr\}\, ,
\end{equation}
where a supersymmetric mass $m$ compatible with the $\Zparity$ parity
has been introduced.
For $F_T\not=0$, the scalar masses are   
given by $|m-\Ft/2\pm n|/R$ and $|m+\Ft/2\pm n|/R$, 
while the fermion sector
has masses
$|m\pm n|/R$.         
This spectrum is  identical to that of Ref.~\cite{pq}
with the identification $m=q_H$ and $F_T=2q_R$.

In conclusion, we have showed that a constant ($y$-independent)
 $W$ yields 
 supersymmetry breaking parametrized
by the   $F$-term of the radion and gives
the same spectrum 
as  supersymmetry broken by boundary conditions.
In particular, we have recovered the mass spectrum of the models
of Refs.~\cite{pq} and \cite{bhn}.
Since supersymmetry is broken 
by the $F$-term of $T$, the Goldstino field corresponds
to  the  fermionic component of $T$ that is 
the fifth-component of the right-handed gravitino, $\Psi_R^5$
(in a different way, this  has also been shown
in Ref.~\cite{goldstino}).

What happens
when  $W$ is localized on the boundary instead of in the bulk?
\footnote{As far as the low-energy effective theory is concerned
(the one describing  physics below the KK masses and 
consisting of  only of the zero modes), it is clear
that there is no difference 
from  the case where $W$ is coming from the bulk as long as
$W/M^3_5\ll 1$ and the zero modes can be considered
 lighter than the KK.}
In this case
the gaugino mass term Eq.~(\ref{majorana}) is
localized on the boundary
and therefore will only affect the even modes
\footnote{See Ref.~\cite{fabio} for the case of the gravitino.}:
\begin{equation}
{\cal L}_{soft}=\frac{\delta(y)}{R}\frac{W}{M^3_5}\lambda_1\lambda_1+
\text{h.c.}\, ,
\label{majoranab}
\end{equation}
for a   canonically normalized $\lambda_1$. 
This term can be thought to arise  from Eq.~(\ref{GaugeAb})    with 
$F_T=2\delta(y)W/M^3_5$. 
The term (\ref{majoranab})
produces a mixing between the KK. To obtain 
the mass spectrum one has to redefine the KK states. 
It is much simpler, however,
to obtain the mass spectrum
by
 solving the 5D equation of motion of the gauginos with the term
Eq.~(\ref{majoranab}) included.
This  has already been done in Ref.~\cite{ahnsw}.
Instead of repeating this here, we will consider this
 scenario of supersymmetry breaking
in a  warped extra dimension.

\subsection{A warped extra dimension}

The interesting situation for a warped extra dimension is when
the breaking of supersymmetry is located on the boundary at $y=\pi$.
In this case the soft masses are  exponentially suppressed,
explaining the hierarchy between the weak and the Planck scale 
\cite{gp,gp2,lsp,dkkls}.
Here we will consider
the effects of triggering a constant 
superpotential
term $W$ on the $y=\pi$ boundary. 
This breaks supersymmetry
 inducing a gaugino mass  given by
\begin{equation}
{\cal L}_{soft}=\frac{\delta(y)}{R}\frac{e^{-Rk\pi }W}
{M^3_5}\lambda_1\lambda_1+\text{h.c.}\, .
\label{majoranabw}
\end{equation}
Eq.~(\ref{majoranabw}) shifts the gaugino mass spectrum
from the gauge-boson one. 
The exact mass eigenvalues $m_n$ of 
the gauginos are derived in the Appendix.
For large supersymmetry breaking, $W\gg M_5^3$, 
the Majorana gaugino masses become independent of $W$: 
\begin{equation}
 \pm \sqrt{\frac{2}{Rk\pi}}ke^{-Rk\pi}\, ,\
\pm \frac{5}{4}\pi ke^{-Rk\pi}\, ,\
\pm \frac{9}{4}\pi ke^{-Rk\pi}\, ,\, ...\, .
\end{equation}
Notice that in 
this limit, $W/M^3_5\rightarrow\infty$,
the gauginos can be combined in  Dirac fermions
and the theory becomes U(1)$_R$ invariant.
This  is exactly the same spectrum 
as the one obtained when supersymmetry is broken by
boundary conditions \cite{gp2}.

For small supersymmetry breaking, $W\ll M_5^3$,
only the zero modes are substantially affected
(the KK masses get  small 
corrections as it is shown in the Appendix).
We can analyze these effects 
by just
considering   the effective theory at energies
below the KK masses with only the zero modes.
For the vector and hypermultiplet sector 
the effective theory 
is given  by Eqs.~(\ref{vector0}) and (\ref{hyper0}).
For the  radion,  the effective lagrangian is 
given by \cite{radion}
\begin{equation}
{\cal L}_{4D}=-\frac{6M^3_5}{k}
\int d^4\mspace{-2mu}\theta \varphi^\dagger\varphi 
(1-e^{-(T+T^\dagger)k\pi})+
\int d^2\mspace{-2mu}\theta 
\varphi^3 \Big[W_0+e^{-3Tk\pi}W+\text{h.c.}\Big]\, ,
\label{radiusaw}
\end{equation}
where $W_0$ and $W$ are  the superpotentials
on the boundary at $y=0$ and 
$y=\pi$ respectively.
$W_0$ has been introduced to cancel the cosmological 
constant as we will see below.
{}From Eq.~(\ref{radiusaw}), we obtain the auxiliary fields
\begin{equation}
F_T=e^{-Rk\pi}\frac{W}{2\pi M^3_5}+\frac{W_0}{2\pi M^3_5}\
,\qquad\qquad  F_\varphi=\frac{k W_0}{2 M^3_5}\, ,
\end{equation}
and the effective radion potential
\begin{equation}
V=\frac{3k}{2M^3_5}\left(e^{-4Rk\pi}|W|^2-|W_0|^2\right)\, .
\end{equation}
Unlike the flat case,
the vanishing of the potential 
is not guaranteed for a constant $W$.
In fact, one must tune 
$|W_0|^2=e^{-4Rk\pi}|W|^2$ to have a zero cosmological 
constant.
In this case, we get
\begin{equation}
F_T=\frac{W}{2\pi M^3_5}e^{-Rk\pi}\
,\qquad\qquad  F_\varphi=
\pi F_Te^{-Rk\pi}\, .
\label{ftwarped}
\end{equation}
We see that, as expected,  
the $F$-term of the radion is exponentially suppressed
and $F_\varphi$, although nonzero, is exponentially smaller
than $F_T$.
Although
this is not a realistic model since $R$ is not stabilized,
it can be considered as a simple example of 
a supersymmetry breaking scenario with  $F_T\sim e^{-Rk\pi}$.
Turning on a nonzero $F_T$
in Eqs.~(\ref{vector0}) and (\ref{hyper0}),
we obtain the supersymmetry breaking masses
\begin{equation}
m_{\lambda_1}=\frac{F_T}{2R}\, , 
\ \ \ \ \ \ \ 
m_\phi=\left|\frac{(1/2-c)k\pi F_T}{2\sinh[(1/2-c)Rk\pi]}\right|\, .
\label{softmasses}
\end{equation}
Notice that $m_\phi$  has its maximum for $c=1/2$ and tends exponentially
to zero   when $c$ deviates  from this value. 

The result of Eq.~(\ref{softmasses})
has an interesting 4D interpretation using
the  AdS/CFT correspondence, that is based on the conjecture
that theories on  AdS$_5$ are dual
to 4D strongly coupled conformal field theories (CFT) 
  in the large $N$ limit \cite{maldacena}.
This correspondence between AdS and CFT theories 
has been also extended 
to the RS set-up,
giving a useful tool to understand the physics of 
this 5D scenario from a 4D point of view.
For example, it has been argued that 
placing the     boundary at $y=0$ 
in the 
AdS$_5$ space corresponds,
in the 4D dual picture, to break explicitly the conformal group
by introducing an ultraviolet cutoff at $k$ and 
by adding new (ultraviolet) degrees of freedom \cite{gu}.
For the case of a 5D theory with gravity and a gauge sector in
the bulk, 
these new degrees of freedom correspond to a graviton and 
a gauge boson coupled to the CFT (it corresponds
to a gauging of the Poincare group and a global symmetry of the CFT).
The
boundary at  $y=\pi$ has a different correspondence in the
4D dual.
It  corresponds
to a spontaneous breaking 
of the conformal group 
at the scale $ke^{-Rk\pi}$. 
The radion is associated to the
 Goldstone boson of the broken conformal symmetry, 
that we will call the dilaton.
This has been recently checked in  different ways in Ref.~\cite{arp}.
Under the conformal transformation, 
$g_{\mu\nu}\rightarrow \Omega^2 g_{\mu\nu}$,
the superfields 
of the effective theory 
of Eqs.~(\ref{vector0})--(\ref{boundtev}) 
must transform 
according to
\begin{eqnarray}
T&\rightarrow& T+\frac{1}{k\pi}\ln\Omega\, ,\nonumber\\
V&\rightarrow& V\, ,\nonumber\\
\Phi &\rightarrow& \Omega^{(c-3/2)}\Phi\, ,\nonumber\\
Q&\rightarrow& Q\, .
\label{trans}
\end{eqnarray}
The lagrangian  of Eqs.~(\ref{vector0}) 
and (\ref{hyper0})
is, however,  not fully invariant under the transformation
Eq.~(\ref{trans}) due, as we said,  
to the  $y=0$ boundary.
In Eq.~(\ref{vector0}) 
the appearance  of $T$ coupled to the vector multiplet
breaks the conformal symmetry.
In the 4D dual picture this corresponds to 
the coupling of  the dilaton  to the gauge vector supermultiplet
at the one-loop level due to the conformal anomaly.
In 5D it appears at 
the tree-level according to the 
duality relation \cite{arp}
$g^2_5b_{CFT}/(8\pi^2)=1/k$.
For the hypermultiplet zero mode, Eq.~(\ref{hyper0}),
the effect of the 
$y=0$ boundary
becomes exponentially suppressed 
for $c\ll 1/2$. In this limit the zero mode
is localized on the  boundary at $y=\pi$ and 
corresponds in the 4D dual picture to a bound state arising from the 
spontaneously broken CFT.

Our scenario where supersymmetry is broken by the $F$-term
of the radion corresponds in the 4D dual 
to break supersymmetry
with the dilaton $F$-term.
This is 
similar to AMSB where the breaking
of supersymmetry is parametrized by the $F$-term of 
the conformal compensator $\varphi$.
In AMSB models
gaugino and scalar masses are generated at the loop level
due to the conformal anomaly.
In our case, the role of $\varphi$ is played by $T$. 
It is also the anomaly (in the 4D dual)  responsible for the 
gaugino masses. 
The scalar mass of $\phi$ tends to zero at tree level
in the conformal limit $c\ll 1/2$, 
but it is generated at the one-loop level, similarly to  AMSB,
due to a wave function renormalization. 
In the opposite limit, 
$c\gg 1/2$, the zero mode of the hypermultiplet is localized
on the $y=0$ boundary and corresponds in the 4D picture
to a new degree of
freedom.
In this case its  tree-level mass 
is also small since its direct coupling to the dilaton (arising from the CFT)
is exponentially suppressed.

\section{Conclusions}

We have presented the 5D action 
of a 
  supersymmetric gauge theory with 
a compact  extra dimension
using  $N=1$ superfields. 
For a flat extra dimension the
action is given in
Eqs.~(\ref{GaugeAb}) and (\ref{eq:hyperS}), while 
for a warped extra dimension as in the RS scenario
this  is given in
Eqs.~(\ref{gaugew}) and (\ref{hyperw}).

We have applied the above results
 to study the breaking of supersymmetry
by the $F$-term of the radion.
We have showed that, 
for  a flat extra dimension,
this type of breaking
 leads to 
the same mass spectrum as 
in  Scherk-Schwarz  models of supersymmetry breaking.
One can therefore
consider our  formulation 
as a superfield description of the SS mechanism.

We have also considered scenarios where supersymmetry
is broken on a boundary of a warped extra dimension.
The spectrum in this case presents certain 
similarities with that
in AMSB
models.

\vskip2cm

\section*{Acknowledgments}

Work  partially supported by the CICYT Research Project
AEN99-0766.

\newpage

\section*{Appendix}

In this Appendix we will derive the mass spectrum  for the gaugino
sector in a theory with a warped extra dimension 
where the breaking of supersymmetry 
induces a 
Majorana gaugino mass on the boundary as in Eq.~(\ref{majoranabw}).

Redefining $\lambda_i \rightarrow
e^{-2R\sigma} \lambda_i$ $i=1,2$ to absorb the spin connection term, the
equations of motion for the gauginos are given by
\begin{align}
    &i e^{R\sigma }\bar{\sigma}^{\mu}\partial_{\mu} \lambda_{2} +
  \frac{1}{R}(\partial_{5} + \frac{1}{2}R \sigma')\bar{\lambda}_1 = 0\, ,
\nonumber\\
    &i e^{R\sigma }\bar{\sigma}^{\mu}\partial_{\mu} \lambda_{1} -
    \frac{1}{R}(\partial_{5} - \frac{1}{2}R \sigma')\bar{\lambda}_2 -
    \frac{1}{2}\frac{W}{M_{5}^{3} R} \delta(y - \pi) 
\bar{\lambda}_{1}= 0\, . 
\label{eq:secnd}
\end{align}
We will solve these equations in the bulk, ignoring boundary effects
which will only play a role when imposing boundary conditions. Looking for
solutions of the form $\lambda_i(x,y)= \sum \lambda^{(n)}(x)
f^{(n)}_{i}(y)$ and using the orthogonality condition of the
modes, Eq.~(\ref{eq:secnd}) leads to
the second order differential equations
\begin{equation}
\Bigl[\frac{1}{R^2}
e^{R\sigma }\partial_{5}(e^{-R\sigma } \partial_{5}) -
\Bigl(\frac{1}{4}\pm \frac{1}{2}\Bigr)k^{2}\Bigr]
f^{(n)}_{1,2} =  e^{2R \sigma } m_n^2 f^{(n)}_{1,2}\;,\label{eq:KKdifeq}
\end{equation}
with solutions
\begin{align}
    f^{(n)}_1(y) & = \frac{e^{R\sigma/2}}{N_n} \Bigl[ J_1
    \left(\frac{m_n}{k} e^{R\sigma }\right) + b_1(m_n)
    Y_1\left(\frac{m_n}{k} e^{R\sigma } \right) \Bigr], \\
    f^{(n)}_2(y) & = \frac{\sigma'}{k}
\frac{e^{R\sigma /2}}{N_n} \Bigl[ J_0
    \left(\frac{m_n}{k} e^{R\sigma }\right) + b_2(m_n)
    Y_0\left(\frac{m_n}{k} e^{R\sigma } \right) \Bigr], 
\end{align}
where $b_i$ and $m_n$ will be determined by the boundary 
conditions, and $N_n$ are normalization constants.
    
Taking into account the $\Zparity$ assignment, 
$f_i^{(n)}$ must fulfill the following conditions on
the $y=0$ boundary:
\begin{align}
    f_2^{(n)} \Bigr\vert_{y=0} & = 0\, ,\nonumber \\
    \biggl(\frac{d}{dy} + \frac{R}{2} \sigma'
     \biggr)f_{1}^{(n)}\Bigr\vert_{y=0} & = 0\, ,
\end{align}
which imply
\begin{equation}
b_1(m_n) = b_2(m_n) = -
\frac{J_0\left(\frac{m_n}{k}\right)}{Y_0\left(\frac{m_n}{k}\right)}
\label{eq:1cnsrt}\, .
\end{equation}
On the other hand, the presence of the Majorana gaugino mass on the
$y=\pi$ boundary in Eq.~(\ref{eq:secnd}) requires the following
condition:
\begin{equation}
    f_{2}^{(n)}(\pi)= 
\frac{1}{2}\frac{W}{2M_{5}^{3} R} f_{1}^{(n)}(\pi)\, .
    \label{eq:2cnsrt}
\end{equation}
Eqs.~(\ref{eq:1cnsrt}) and (\ref{eq:2cnsrt}) yield
\begin{equation}
2\Bigl[J_0 \left(\frac{m_n}{k} e^{Rk\pi } \right) -
\frac{J_0\left(\frac{m_n}{k} \right)}{Y_0\left(\frac{m_n}{k} \right)}
Y_0\left(\frac{m_n}{k} e^{Rk\pi } \right) \Bigr] = \frac{W}{2M_{5}^{2} R}
\Bigl[J_1 \left(\frac{m_n}{k} e^{Rk\pi } \right) -
\frac{J_0\left(\frac{m_n}{k} \right)}{Y_0\left(\frac{m_n}{k} \right)}
Y_1\left(\frac{m_n}{k} e^{Rk\pi } \right) \Bigr]\, ,
\label{eq:Cond}
\end{equation}
that determines the mass spectrum.
Let us look for solutions of Eq.~(\ref{eq:Cond})
in the limit $kR \gg 1$. 
For the lightest modes
we can take
the limit  
 $m_n e^{R k \pi} /k \equiv \epsilon\ll 1$ 
in  Eq.~(\ref{eq:Cond})
that becomes
\begin{equation}
2 = \frac{W}{2 M_{5}^{3}} \left[\frac{\epsilon}{2} - \frac{1}{\pi k R}
    \frac{1}{\epsilon} \right]\, .
\label{eq:zerom}
\end{equation}
If the breaking of supersymmetry is small,  $W/M_{5}^{3} \ll 1$,
the only solution to Eq.~(\ref{eq:zerom}) 
fulfilling $\epsilon\ll 1$ 
is given by
\begin{equation}
 m \simeq \frac{W}{4 M_{5}^{3} \pi R} e^{-Rk\pi}\, .
\end{equation}
This corresponds to the zero mode mass.
In the strong supersymmetry breaking case, 
$W/M_{5}^{3} \gg 1$,
there are two solutions to Eq.~(\ref{eq:zerom}) given by
\begin{equation}
 m \simeq \pm\sqrt{\frac{2}{\pi k R}} k e^{-Rk\pi}\; .
\end{equation}
On the other hand, the masses of the  heavier KK modes are easily
obtained from Eq.~(\ref{eq:Cond})
 in the limit $\epsilon> 1$ that becomes
\begin{equation}
 \frac{J_0 \left(\frac{m_n}{k} e^{Rk\pi } \right)}{J_1
  \left(\frac{m_n}{k} e^{Rk\pi } \right)} = \frac{W}{4 M_{5}^{3}}\, .
\end{equation}
If the breaking of supersymmetry is weak, 
$W/M_{5}^{3} \ll 1$,
the Majorana mass spectrum is
approximately given by
 \begin{equation}
m_n \simeq 
\left(n + \frac{3}{4} \pm \frac{W}{4\pi M_{5}^{3}}\right) \pi k
   e^{-Rk\pi }\, , \ \ \ n=1,2,...,
\end{equation}
whereas in the strong supersymmetry breaking limit, 
$W/M_{5}^{3} \gg 1$, we have
\begin{equation}
m_n \simeq \left(n + \frac{1}{4} 
\pm \frac{4 M_{5}^{3}}{\pi W}\right) \pi k
   e^{-Rk\pi }\; .
\end{equation}

\end{document}